\documentstyle[epsbox]{article}

\date{}
\input epsf
\title{\bf Confinement and Topological Charge in~the~Abelian~Gauge of QCD}
\author{\underline{H. Suganuma},
S. Sasaki, H. Ichie, F. Araki and O. Miyamura$^{\rm a}$,\\
\\
\small \it Rearch Center for Nuclear Physics (RCNP), Osaka University,\\
\small \it Mihogaoka 10-1, Ibaraki, Osaka 567, Japan\\
\\
\small $^{\rm a}$\it Department of Physics, Hiroshima University,\\
\small \it Kagamiyama 1-3, Higashi-Hiroshima 739, Japan}
\begin{document}

\maketitle

\begin{abstract}
We study the relation between instantons and monopoles in the abelian gauge. 
First, we investigate the monopole in the multi-instanton solution 
in the continuum Yang-Mills theory using the Polyakov gauge.
At a large instanton density, the monopole trajectory becomes 
highly complicated, which can be regarded as a signal of monopole condensation. 
Second, we study instantons and monopoles in the SU(2) lattice gauge theory 
both in the maximally abelian (MA) gauge and in the Polyakov gauge. 
Using the $16^3 \times 4$ lattice, we find monopole dominance for instantons  
in the confinement phase even at finite temperatures.
A linear-type correlation  is found between the total monopole-loop length  
and the integral of the absolute value of the topological density 
(the total number of instantons and anti-instantons) in the MA gauge.
We conjecture that instantons enhance the monopole-loop length 
and promote monopole condensation.

\end{abstract}

\section{Analytical Study for Monopole Trajectory in the Multi-instanton Solution}

As 't~Hooft pointed out, a nonabelian gauge theory is reduced into 
an abelian gauge theory with monopoles by the abelian gauge fixing [1,2].
Recent lattice studies suggest abelian dominance  
and relevant roles of monopole condensation [3]
for the nonperturbative phenomena: confinement [4], 
chiral symmetry breaking [5] and instantons [5-8].
In the abelian gauge, 
unit-charge monopoles appear from the hedgehog-like gauge configuration 
according to the nontrivial homotopy group, 
$\pi_{2}\{{\rm SU}(N_c)/{\rm U}(1)^{N_c-1}\}=Z^{N_c-1}_\infty$ [2].

On the other hand, the instanton is another relevant topological object
in the nonabelian gauge manifold ($\pi_{3}({\rm SU}(N_c))$ =$Z_\infty$).
In the abelian-dominant system, 
the instanton seems to lose the topological basis for its existence, 
and hence it seems unable to survive in the abelian manifold [7-9].
However, even in the abelian gauge, nonabelian components remain
relatively large around the topological defect, {\it i.e.} monopoles, and 
therefore instantons are expected to survive only around the monopole world lines
in the abelian-dominant system [7-9].

We have pointed out such a close relation between instantons and monopoles, and have demonstrated it 
in the continuum Yang-Mills theory using the Polyakov gauge, where $A_4(x)$ is diagonalized [7-9].
We summarize  our previous analytical works as follows [7-9]. \\
(1) Each instanton center is penetrated by a monopole world line in the Polyakov gauge, 
because $A_4(x)$ takes a hedgehog configuration near the instanton center. 
In other words, instantons only live along the monopole trajectory. \\
(2) Even at the classical level, the monopole trajectory is unstable against a small fluctuation 
of the location or the size of instantons,  although it is relatively stable inside the instanton profile. \\
(3) In the two-instanton solution, a loop or folded structure appears in the monopole trajectory
depending on the instanton location and size.\\
(4) In the multi-instanton solution, monopole trajectories become very unstable and complicated. \\  
(5) At a high temperature, the monopole trajectories are drastically changed, and become simple lines along 
the temporal direction.
 
To begin with, we study the monopole trajectory in the multi-instanton system 
in terms of the topological charge density as a gauge invariant quantity.
We show in Fig.1 an example of the monopole trajectory in the Polyakov gauge 
in the multi-instanton system, where all instantons are randomly 
put on the $zt$-plane for simplicity [8].
The contour denotes the magnitude of the topological density.
Each instanton attaches the monopole trajectory.
As the instanton density increases, the monopole trajectory tends to be  highly complicated and very long, 
which can be regarded as a signal of monopole condensation [3,10].
As a remarkable feature in the Polyakov gauge,
the monopole favors the high topological density region, ``mountain'': 
each monopole trajectory walks crossing tops of the mountain [8].
On the other hand, anti-monopole with the opposite color-magnetic
charge favors the low topological density region, ``valley".
Thus, the strong local correlation is found between the instanton and
the monopole trajectory [6-9].

\begin{figure}[htb]
\begin{center}
\epsfile{file=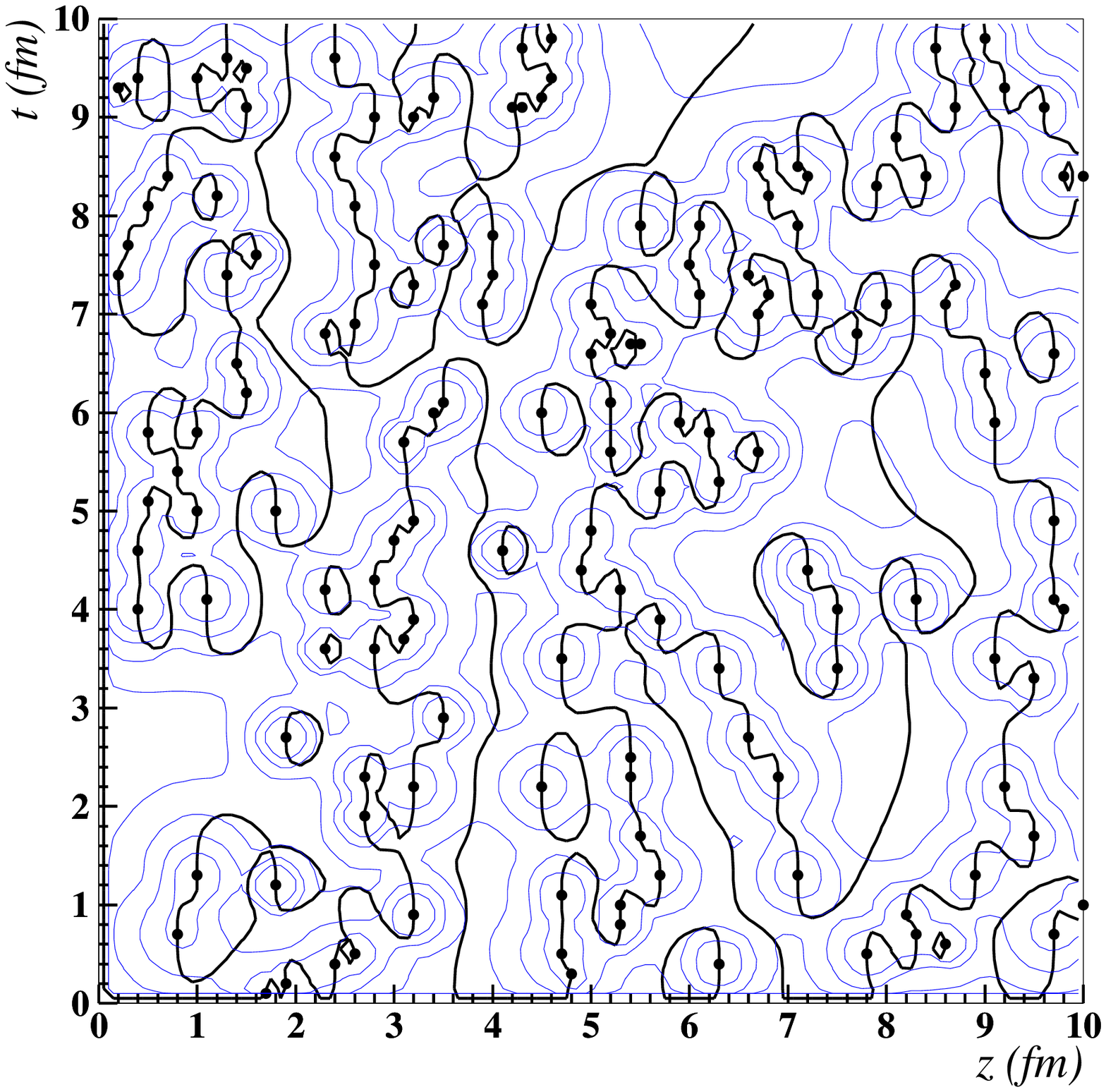,height=5.8cm}
\end{center}

{
\small
\noindent
Fig.1: The monopole trajectory in the Polyakov gauge in the multi-instanton solution, 
where 150 instantons are put on the $zt$ plane.
The contour denotes the magnitude of the topological density.
}
\end{figure}

\section{Instanton and Monopole at Finite Temperature on SU(2) Lattice}

Next, we study the correlation between instantons and monopoles 
in the maximally abelian (MA) gauge and in the Polyakov gauge
using the Monte Carlo simulation in the SU(2) lattice gauge theory [5-8].
The SU(2) link variable can be separated into the monopole-dominating (singular) part
and the photon-dominating (regular) part [4,5,7-8].
Using the cooling method, we measure the topological quantities ($Q$ and $I_Q$) 
in the monopole and photon sectors as well as in the ordinary SU(2) sector. 
Here, $I_Q \equiv \int d^4x |{\rm tr}(G_{\mu\nu} \tilde G_{\mu\nu})|$ 
corresponds to the total number $N_{\rm tot}$ of instantons and anti-instantons.

(1) On the $16^4$ lattice with $\beta=2.4$, we find that instantons exist only in the monopole part 
both in the MA and Polyakov gauges, which means monopole dominance for the topological charge [5,7,8].
Hence, we can expect monopole dominance for the U$_{\rm A}$(1) anomaly and the $\eta'$ mass.

(2) We study the finite-temperature system using the
$16^3 \times 4$ lattice with various $\beta$ around $\beta_c \simeq 2.3$ [8].
We show in Fig.2 the correlation between $I_Q({\rm SU(2)})$ and $I_Q({\rm Ds})$, which are measured 
in the SU(2) and monopole sectors, respectively, after 50 cooling sweeps.
The monopole part holds the dominant topological charge in the full SU(2) gauge configuration.
On the other hand, $I_Q(Ph)$, measured in the photon part, vanishes quickly by several cooling sweeps. 
Thus, monopole dominance for the instanton is found also in the confinement phase 
even at finite temperatures [8].

(3) Near the critical temperature $\beta_c \simeq 2.3$, a large reduction of $I_Q$ is observed.
In the deconfinement phase, $I_Q$ vanishes quickly by several cooling sweeps, 
which means the absence of the instanton, in the SU(2) and monopole sectors 
as well as in the photon sector [8].
Therefore, the gauge configuration becomes similar to the photon part 
in the deconfinement phase.

\begin{figure}[ht]
\begin{center}
\epsfile{file=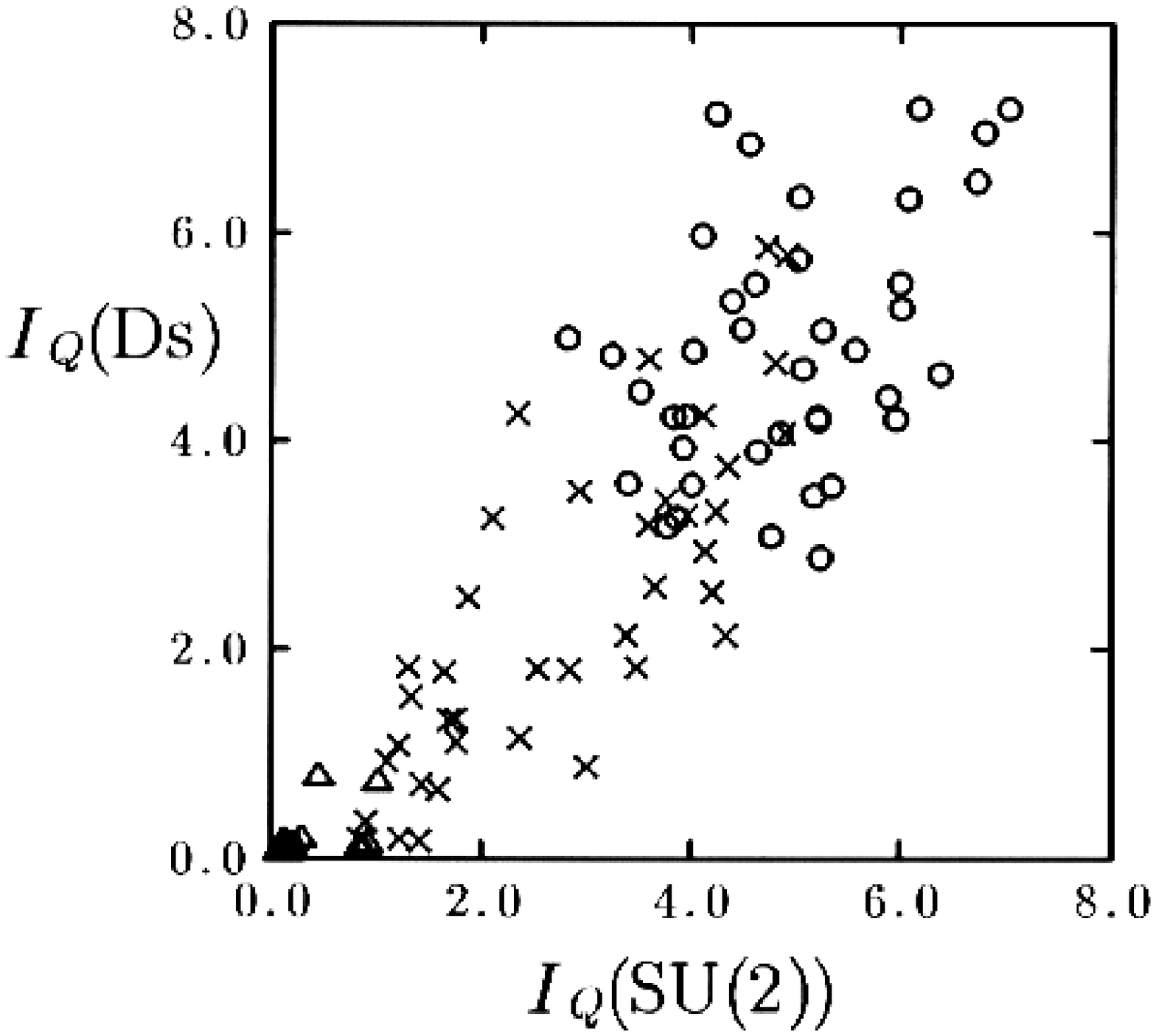,height=5.8cm}
\end{center}

{
\small
\noindent 
Fig.2 : Correlation between $I_Q({\rm SU(2)})$ and $I_Q({\rm Ds})$ at $\beta$=2.2 ($\circ$), 
2.3 ($\times$), 2.35 ($\bigtriangleup$) after 50 cooling sweeps. 
}
\end{figure}
Thus, monopole dominance for instantons is found in the confinement phase even at finite temperatures, and the monopole part includes a dominant amount of instantons 
as well as the monopole current [5,7].  
At the deconfinement phase transition, both the instanton density and the monopole current are 
rapidly reduced, and the QCD-vacuum becomes trivial in terms of the topological nontriviality.
Because of such strong correlation between instantons and monopoles, 
monopole dominance for the nonperturbative QCD may be interpreted as  
instanton dominance.

\section{Lattice Study for Monopole Trajectory and Instantons}

Finally, we study the correlation between the instanton number and the monopole loop length.
Our analytical studies suggest appearance of a highly complicated  monopole trajectory  
in the multi-instanton system even at the classical level [7-9]. 
Further monopole clustering would be brought by quantum effects. 
We conjecture that the existence of instantons promotes monopole condensation [7-9],
which is characterized by a long complicated monopole trajectory covering over ${\bf R}^4$ 
in the similar argument for the Kosterlitz-Thouless transition [10] 
and is observed also in the lattice QCD simulation [3,4,8].
 
To clarify the role of instantons on monopole condensation, we study the SU(2) lattice gauge theory 
for the total monopole-loop length $L$ and the integral of the absolute value of the topological density 
$I_Q$, which corresponds to the total number $N_{\rm tot}$ of instantons and anti-instantons.
We plot in Fig. 3 the correlation between $I_Q$ and $L$ in the MA gauge after 10 cooling sweeps  
on the $16^3 \times 4 $ lattice with various $\beta$ .
A linear-type correlation is clearly found between $I_Q$ and $L$. 
Hence, the monopole-loop length would be largely enhanced in the dense instanton system. 
\begin{figure}[htb]
\begin{center}
\epsfile{file=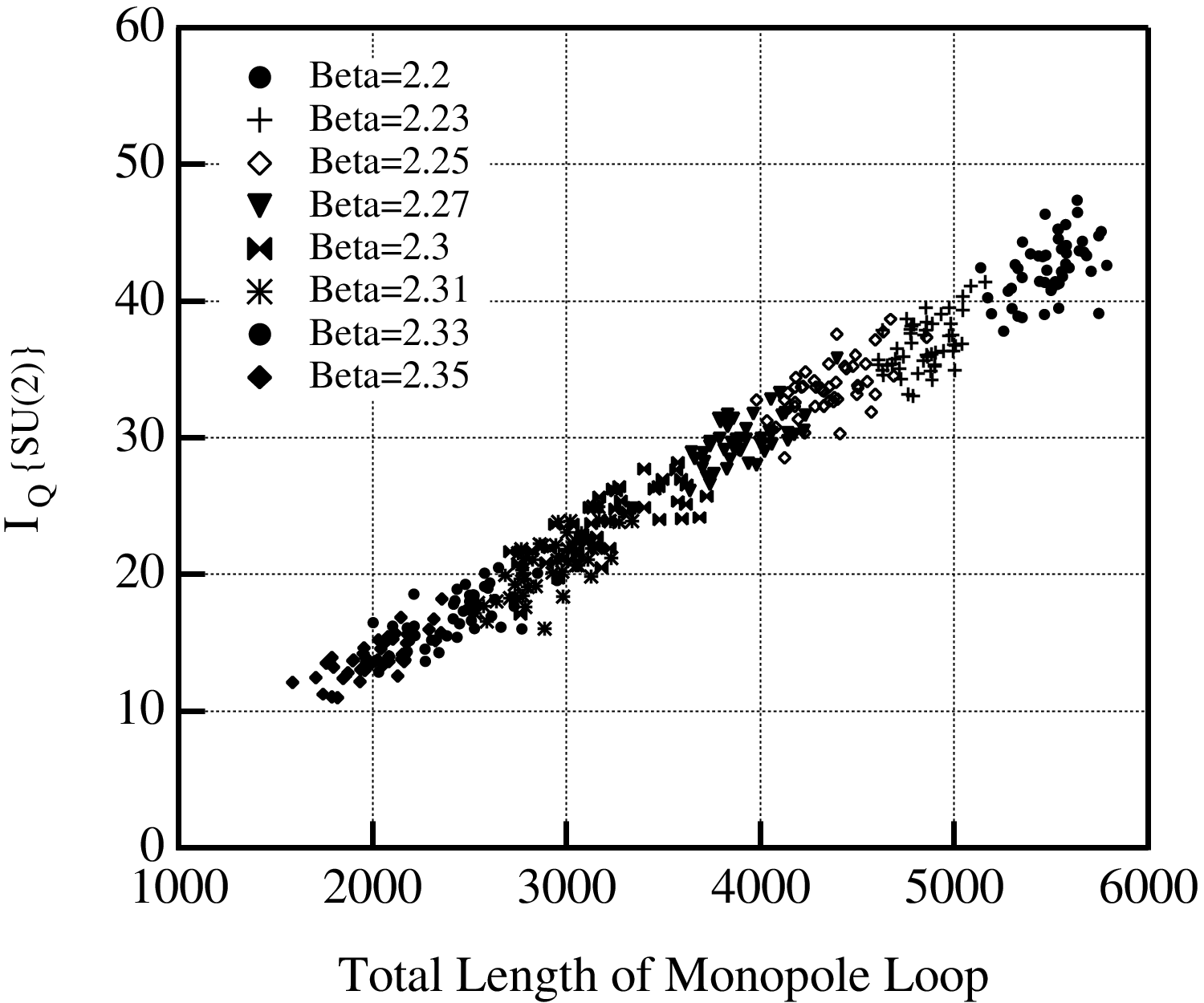,height=5.8cm}
\end{center}

{
\small
\noindent
Fig.3 : Correlation between the total monopole-loop length $L$ and $I_Q$
(the total number of instantons and anti-instantons) in the MA gauge.
We plot the data at 10 cooling sweep on $16^3 \times 4$ lattice 
with various $\beta$. 
}
\end{figure}

From the above results, we propose the following conjecture.
Each instanton accompanies a small monopole loop nearby, 
whose length would be proportional to the instanton size [11-14].
When $N_{\rm tot}$ is large enough, these monopole loops overlap, and there appears
a very long monopole trajectory, which bonds neighboring instantons [8,11].
Such a monopole clustering leads to monopole condensation and color confinement [10].
Thus, instantons would play a relevant role on color confinement by providing a source of 
the monopole clustering [7-9,11].




\clearpage

\end{document}